\documentclass[a4paper,12pt]{article}
\usepackage[pctex32]{graphics}

\begin{document}
\topmargin 0pt
\oddsidemargin 0mm

\newcommand{\alp}{\alpha}
\newcommand{\bta}{\beta}
\newcommand{\gmm}{\gamma}
\newcommand{\del}{\delta}
\newcommand{\omg}{\omega}
\newcommand{\sgm}{\sigma}
\newcommand{\lmd}{\lambda}
\newcommand{\tha}{\theta}
\newcommand{\vph}{\varphi}
\newcommand{\Alp}{\Alpha}
\newcommand{\Bta}{\Beta}
\newcommand{\Gmm}{\Gamma}
\newcommand{\Del}{\Delta}
\newcommand{\Omg}{\Omega}
\newcommand{\Sgm}{\Sigma}
\newcommand{\Lmd}{\Lambda}
\newcommand{\Tha}{\Theta}
\newcommand{\half}{\frac{1}{2}}
\newcommand{\rnd}{\partial}
\newcommand{\nab}{\nabla}

\newcommand{\beqn}{\begin{eqnarray}}
\newcommand{\eeqn}{\end{eqnarray}}
\newcommand{\be}{\begin{equation}}
\newcommand{\ee}{\end{equation}}

\begin{titlepage}

\vspace{5mm}
\begin{center}
{\Large \bf Entropic force in the presence of  black hole }
\vspace{12mm}

{\large   Yun Soo Myung \footnote{e-mail
 address: ysmyung@inje.ac.kr}}
 \\
\vspace{10mm} {\em  Institute of Basic Science and School of
Computer Aided Science, Inje University, Gimhae 621-749, Republic of
Korea}

\end{center}

\vspace{5mm} \centerline{{\bf{Abstract}}}
 \vspace{5mm}
We derive the entropic force in the presence of the Schwarzschild
black hole by using the local equipartition rule and holographic
principle. On the other hand, when using the Tolman temperature, one
does not arrive at the Newtonian force law.

\end{titlepage}
\newpage
\renewcommand{\thefootnote}{\arabic{footnote}}
\setcounter{footnote}{0} \setcounter{page}{2}

\section{Introduction}
Recently, Verlinde has proposed the Newtonian law of gravity as an
entropic force by using the holographic principle and the
equipartition rule~\cite{Ver}. After released his work,  the
dynamics of apparent horizon in the Friedmann-Robertson-Walker
universe~\cite{SG}, the Friedmann equations~\cite{Pad,CCO}, the
connection in the loop quantum gravity~\cite{Smo}, holographic
actions for black hole entropy~\cite{CM}, and application to
holographic dark energy~\cite{LWh} were considered from the entropic
force. Furthermore,  cosmological implications were reported in
\cite{Gao,ZGZ,Wang,WLW,LW,LKL}, and the extension to Coulomb
force~\cite{WangT} and the symmetry aspect of entropy
force~\cite{Zhao} were investigated.

Explicitly, when a test particle with mass $m$ is located near a
holographic screen with distance $\Delta x$, the change of entropy
on the holographic screen takes the form \be \label{eq1} \Delta S=
2\pi m \Delta x \ee in the natural units of $\hbar=c=k_B=1$ and
$G=l^2_{pl}$. Considering that the entropy of a system depends on
the distance $\Delta x$, an entropic force $F$ could be arisen from
the thermodynamical conjugate of the distance \be \label{eq2} F
\Delta x=T \Delta S \ee which is regarded as an indication that the
first law of thermodynamics is satisfied on the holographic screen.
Plugging (\ref{eq1}) into (\ref{eq2}) leads to a connection between
entropic force and temperature \be \label{eq3} F=2\pi m T. \ee Let
us assume that the energy $E$ is distributed on the spherical shape
of holographic screen with radius $R$, where the mass $M$ is located
at the origin of coordinate. Then, we introduce the equipartition
rule, the equality of  energy and mass, and the holographic
principle, respectively,  as \be \label{eq4} E=\frac{1}{2 }N
T,~~E=M,~~N=\frac{A}{G}=4S \ee with the area of the holographic
screen $A=4\pi R^2$. These are combined to  provide the temperature
on the screen \be \label{eq5} T=\frac{GM}{2\pi R^2}. \ee
Substituting (\ref{eq5}) into (\ref{eq3}), one obtains the Newtonian
force law as  the entropic force\be \label{eq6} F=\frac{G m M}{R^2}.
\ee

In this work, we will derive the entropic force by replacing the
mass $M$ by the Schwarzschild black hole with mass $M$ as \be
\label{eq7} ds^2_{Sch}=g_{\mu\nu}dx^\mu
dx^\nu=-\Big(1-\frac{2GM}{r}\Big)dt^2+\frac{dr^2}{\Big(1-\frac{2GM}{r}\Big)}+r^2d\Omega^2_2.
\ee Here the even horizon (EH) is located at $r=r_{EH}=2GM$ whose
horizon area is given by $A_{EH}=4\pi r_{EH}^2$.

\section{Entropic force in the presence of black hole}
It is well known that in the presence of the black hole, we may
introduce the local temperature (Tolman temperature) and the local
energy on the holographic screen. Now let us first  introduce the
Tolman redshift transformation on the black hole system~\cite{GPP2}.
In general, the local temperature observed by an observer at
$r>r_{EH}$ outside the Schwarzschild black hole  is defined
by~\cite{York} \be \label{eq8} T_L(r)=
\frac{T_{\infty}}{\sqrt{-g_{tt}}}=\frac{1}{8\pi G
M}\frac{1}{\sqrt{1-\frac{2GM}{r}}} \ee where \be \label{eq9}
T_\infty=\frac{1}{8\pi G M}=\frac{1}{4 \pi r_{EH}}\equiv T_H \ee is
the Hawking temperature $T_H$  measured at infinity and the
denominator of $\sqrt{-g_{tt}}$ is the red-shifted factor. On the
holographic screen at $r=r_{EH}+l_{pl}^2/r_{EH}$ near the event
horizon, this local temperature is given by $T=1/8 \pi l_{pl}$ which
is independent of the black hole mass $M$~\cite{Myungent}. On the
other hand, for $r \gg r_{EH}$, it reduces to the Hawking
temperature $T_H$.

Similarly, the local energy is given by \be \label{loen} E_L(r)=
\frac{E_{\infty}}{\sqrt{-g_{tt}}} \ee
 where  the energy observed  at infinity is the ADM mass $M$
 \be
 \label{loeni}
 E_{\infty}=M=\frac{r_{EH}}{2G}.\ee
 Eq.(\ref{loen})  states clearly the UV/IR scaling transformation (Tolman redshift transformation) of the energy  between the bulk and
 the holographic screen.
  On the holographic screen near the
horizon, we have $ E_L(r_{EH}+l_{pl}^2/r_{EH}) \propto
A_{EH}$~\cite{MSeo}.

 However, it is very important to note that
there is no difference between the local black hole  entropy $S_{L}$
near the horizon and the entropy $S_{\infty}$ at infinity: \be
\label{entrop} S_{L}=\pi r_{EH}^2=S_\infty,\ee which is surely the
Bekenstein-Hawking entropy for the Schwarzschild black hole. That
is, the black hole entropy is invariant under the UV/IR
transformation.

Now we are in a position to  derive the entropic force on the
holographic screen in the presence of Schwarzschild black hole. To
this end, we propose the local equipartition rule \be \label{eq11}
E_L(r)=2S_{S}(r)T_S(r), \ee where the entropy  $S_S$ is defined on
the holographic screen located at $r
>r_{EH}$ \be \label{eq12} S_{S}(r)= \frac{\pi r^2}{G}. \ee
 Then, the temperature on the screen is given by
 \be
 \label{eq13}
 T_S(r)=\frac{GM}{2\pi r^2 \sqrt{-g_{tt}}}. \ee
Plugging (\ref{eq13}) into (\ref{eq3}), we obtain the entropic
 force as
 \be
 \label{eq14}
 F_{BH}= 2\pi m T_S(r)= \frac{1}{\sqrt{1-\frac{r_{EH}}{r}}} \frac{GmM}{ r^2} \ee
 which shows that  the mass $m$  feels an infinitely tidal force  in the limit of $r \to r_{EH}$,
 while it takes  the Newtonian force law at  the large distance of  $r\gg
 r_{EH}$. This is our main result.

 On the other hand, the temperature $T_S(r)$  on the screen becomes
  the local temperature $T_L(r)$ in (\ref{eq8}) when using the Bekenstein-Hawking entropy (\ref{entrop})
 for (\ref{eq11}).  In this case, Eq.(\ref{eq3}) takes a different form
 \be
 F=2\pi m T_L(r)= \frac{m}{4GM} \frac{1}{\sqrt{1-\frac{r_{EH}}{r}}},\ee
 which is nothing to do with the entropic force.

\section{Discussions}
We have obtained the entropic force in the presence of the
Schwarzschild black hole by using the local equipartition rule and
the holographic principle on the screen. In this case, the local
energy $E_L$  together with the entropy $S_S$ on the screen  was
mainly used to derive the entropic force. When using the local
(Tolman) temperature and the Bekenstein-Hawking entropy for the
Schwarzschild black hole, we have failed to arrive at the Newtonian
force law. At this stage, it is not clear why the latter approach is
not suitable for deriving the entropic force.

\section*{Acknowledgment}
The author  thanks Rong-Gen Cai for helpful discussions on entropic
force. This work  was supported by Basic Science Research Program
through the National  Research  Foundation (NRF)  of  Korea funded
by the Ministry of Education, Science and Technology (2009-0086861)
and


\begin{thebibliography}{99}

\bibitem{Ver}
  E.~P.~Verlinde,
  arXiv:1001.0785 [hep-th].


\bibitem{SG}
  F.~W.~Shu and Y.~Gong,
  arXiv:1001.3237 [gr-qc].

\bibitem{Pad}
  T.~Padmanabhan,
  arXiv:1001.3380 [gr-qc].



\bibitem{CCO}
  R.~G.~Cai, L.~M.~Cao and N.~Ohta,
  arXiv:1001.3470 [hep-th].

\bibitem{Smo}
  L.~Smolin,
  arXiv:1001.3668 [gr-qc].


\bibitem{CM}
  F.~Caravelli and L.~Modesto,
  arXiv:1001.4364 [gr-qc].

\bibitem{LWh}
  M.~Li and Y.~Wang,
  arXiv:1001.4466 [hep-th].


\bibitem{Gao}
  C.~Gao,
  arXiv:1001.4585 [hep-th].

\bibitem{ZGZ}
  Y.~Zhang, Y.~g.~Gong and Z.~H.~Zhu,
  arXiv:1001.4677 [hep-th].



\bibitem{Wang}
  Y.~Wang,
  arXiv:1001.4786 [hep-th].


\bibitem{WLW}
  S.~W.~Wei, Y.~X.~Liu and Y.~Q.~Wang,
  arXiv:1001.5238 [hep-th].


\bibitem{LW}
  Y.~Ling and J.~P.~Wu,
  arXiv:1001.5324 [hep-th].

\bibitem{LKL}
  J.~W.~Lee, H.~C.~Kim and J.~Lee,
  arXiv:1001.5445 [hep-th].


  \bibitem{WangT}
  T.~Wang,
  arXiv:1001.4965 [hep-th].

\bibitem{Zhao}
  L.~Zhao,
  arXiv:1002.0488 [hep-th].

\bibitem{GPP2}
  G.~W.~Gibbons, M.~J.~Perry and C.~N.~Pope,
  Phys.\ Rev.\ D {\bf 72}, 084028 (2005)
  [hep-th/0506233].


\bibitem{York}
  J.~W.~York,
  Phys.\ Rev.\ D {\bf 33} (1986) 2092.
\bibitem{Myungent}
  Y.~S.~Myung,
  Phys.\ Lett.\  B {\bf 636} (2006) 324
  [arXiv:gr-qc/0511104].

\bibitem{MSeo}
  Y.~S.~Myung and M.~G.~Seo,
  Phys.\ Lett.\  B {\bf 671} (2009) 435
  [arXiv:0803.2913 [gr-qc]].








\end{thebibliography}
\end{document}